\documentclass[10pt,preprint2,a4paper]{aastex}
\usepackage{graphics,epsf}
\usepackage{amsmath}                
\usepackage{amsfonts}               
\usepackage{amssymb}                
\usepackage{epsfig}                 

\usepackage[left=2.3cm,top=2.3cm,right=2.2cm,bottom=2.0cm,nohead,nofoot]{geometry}

\def \K{~\rm{K}}

\def \AU{~\rm{AU}}
\def \erg{~\rm{erg}}
\def \yrs{~\rm{yrs}}
\def \yr{~\rm{yr}}

\def \kpc{~\rm{kpc}}

\def \mum{~\rm{\mu m}}

\begin{document}

\title{THE OUTCOME OF THE PROTOPLANETARY DISK OF VERY MASSIVE STARS}

\author{Amit Kashi\altaffilmark{1} and Noam Soker\altaffilmark{1}}

\altaffiltext{1}{Department of Physics, Technion $-$ Israel Institute of
Technology, Haifa 32000 Israel; kashia@physics.technion.ac.il;
soker@physics.technion.ac.il.}
\setlength{\columnsep}{1.0cm}
\small

\begin{abstract}
We suggest that planets, brown dwarfs, and even low mass stars can be formed by
fragmentation of protoplanetary disks around very massive stars ($M \gtrsim 100 \rm{M_{\odot}}$).
We discuss how fragmentation conditions make the formation of very massive planetary systems
around very massive stars favorable.
Such planetary systems are likely to be composed of brown dwarfs and
low mass stars of $\sim 0.1-0.3 \rm{M_{\odot}}$,
at orbital separations of $\sim {\rm few} \times 100 - 10^4 \AU$.
In particular, scaling from solar-like stars suggests that hundreds of Mercury-like planets
might orbit very massive stars at $\sim 10^3 \AU$, where conditions might favor liquid water.
Such fragmentation objects can be excellent targets for the James Webb Space Telescope
and other large telescopes working in the IR bands.
We predict that deep observations of very massive stars would reveal
these fragmentation objects, orbiting in the same orbital plane
in cases where there are more than one object.
\end{abstract}

\keywords{protoplanetary disks --- planets and satellites: formation --- stars: formation --- stars: massive
--- stars: low-mass --- (stars:) brown dwarfs}

\section{INTRODUCTION}
\label{sec:intro}

Planets were found around objects which had been considered not to be able to support planets,
such as around the pulsars PSR1829–10 (Bailes et al. 1991) and PSR1257+12 (Wolszczan \& Frail 1992),
where they are believed to be formed after the supernova explosion.
In other cases planets can survive the post main sequence evolution of their parent stars.
Examples include a planetary mass companion in orbit around V391 Pegasi (Silvotti et al. 2007),
and the close planet around HD149382 (Geier et al. 2009), both are extreme horizontal branch stars.
These planets survived the red giant branch phase of their parent star.
There was also a tentative detection of a planet around a white dwarf
(e.g., Mullally et al. 2008).
Theoretical models even predict formation of second generation planets
in disks around MS stars accreting mass from their AGB companions (Perets 2010).
In this paper we are studying the possibility of finding planets, brown dwarfs (BDs) and low mass stars orbiting in the
same plane around very massive stars (VMS, $M_* \ga 100 \rm{M_\odot}$).

For planets to exist around VMSs their orbital distance must be very large.
A relevant and famous extrasolar planet was observed around the $\sim 2 \rm{M_{\odot}}$
star Fomalhaut (Kalas et al. 2008).
The orbital distance of the observed planet Fomalhaut b is $\sim 115\AU$,
and its mass was constrained to be $\lesssim 3 \rm{M_J}$ (Chiang et al. 2009).
Fomalhaut b teaches us that planets can be formed at large distances from their parent star.

According to common view, gas giant planets can be formed through two main channels
(e.g., Boley 2009; Dodson-Robinson et al. 2009): accretion of planetesimals into a core, followed by gas accretion
(e.g., Pollack et al. 1996; Kenyon \& Bromley 2009; Brauer et al. 2008), and instabilities in the protoplanetary
disk (PPD) that causes fragmentation (e.g., Cameron 1978; Mayer et al. 2002; Boss 1997, 2006;
Brauer et al. 2008).
Dodson-Robinson et al. (2009) showed that giant planets in wide orbits
of $r \gtrsim 35 \AU$ form by instabilities in the PPD.
Stamatellos \& Whitworth (2008) found that fragmentation does not occur below $r \lesssim 40 \AU$,
a result in agreement with Dodson-Robinson et al. (2009), considering the somewhat different
parameters used.

Rice et al. (2003) found using hydrodynamic simulations and N-body orbit integrations that
when a $0.1 \rm{M_{\odot}}$ PPD around a $1 \rm{M_{\odot}}$ star fragments,
the most massive object to be formed is a $0.0075 \rm{M_{\odot}}$ BD.
As the simulation of Rice et al. (2003) is scale-free, it immediately implies that a PPD
around a $100 \rm{M_{\odot}}$ star is expected to fragment into a low mass star of up to $0.75\rm{M_{\odot}}$.
A more detailed estimate carried in section \ref{sec:frag} gives a somewhat lower mass.

In this paper we examine the possibility that planets, BDs,
and in particular low mass stars can be formed around a VMS.
We start with two basic assumptions:
(1) Very massive single and binary stars posses circumstellar and circumbinary disks
similar to those of low mass stars.
(2) We can scale the constraints and conditions from low mass stars to VMSs.
This assumption is based in part on the results of Kratter \& Matzner (2006) who studied massive PPDs.
Although they did not study explicitly the regime explored here,
some of their parameters overlap with those studied by us.
Following these basic assumptions, in section \ref{sec:frag} we derive scaling relations, and
show that fragmentation around VMSs can occur, and produce fragmentation objects
as massive as low mass stars.
In section \ref{sec:obs} we discuss the observational signature of the proposed fragmentation objects.
We summarize our results in section \ref{sec:summary}.

\section{DISK FRAGMENTATION AROUND A VERY MASSIVE STAR}
\label{sec:frag}

We wish to determine the properties of a fragmented protoplanetary disk (PPD)
around a VMS; we scale by $M_* = 100 \rm{M_{\odot}}$.
The PPD is assumed to be vertically optically thick (i.e., perpendicular to the disk plane)
at the distance where it fragments (Kratter et al. 2010).
The surface density profile of the PPD is assumed to have the same form as in low mass stars
\begin{equation}
\Sigma(r) = \Sigma_0 \left(\frac{r}{1 \AU}\right)^{-p} \rm{~g~cm^{-2}},
\label{eq:sigma1}
\end{equation}
where $\Sigma_0$ and $p$ are constants (Weidenschilling 1977; Hayashi 1981; Nero \& Bjorkman 2009).
The power $p$ varies from one PPD to the other in the range $p=0.5-3$ (e.g., Nero \& Bjorkman 2009),
but more commonly in the range $p=1-3/2$.
For early stages of PPD evolution, where steady accretion from the PPD to the star takes place,
an analytical solution to the evolutionary equation of the PPD (e.g., Frank, King \& Raine 2002)
gives $p=1$.
At later stages the accretion to the star decreases and the value of $p$ increases, until
accretion stops and then an analytical solution to the evolutionary equation of the PPD
gives $p=3/2$.
There are hints that practically, for more massive stars a smaller value of $p$ should be used.
For the solar protoplanetary nebula a value of $p=3/2$ is usually used
(Weidenschilling 1977; Hayashi 1981),
but other values, such as $p=2.168$ (Desch 2007), were also suggested.
The models of Vaidya et al. (2009) for PPDs around $10-37\rm{M_{\odot}}$
stars were best fitted with $p=1.1$.
The stellar disk structure analysis of Andrews \& Williams (2007) based on high resolution
submillimeter continuum survey of circumstellar disks in the Taurus-Auriga and Ophiuchus-Scorpius star
formation regions, also supports a value of $p \simeq 1$.
For the solar PPD, $\Sigma_0$ is taken to be in the range of $1700 \rm{~g~cm^{-2}}$ (Hayashi 1981)
to $4200 \rm{~g~cm^{-2}}$ (Weidenschilling 1977).
For the disk around Fomalhaut the model of Nero \& Bjorkman (2009) implies a value of
$\Sigma_0 \simeq 10^3 \rm{~g~cm^{-2}}$.

Our goal is to determine the range of the orbital distances $r_p$ where fragmentation around
VMSs occurs, and the outer radius of the PPD, $R_d$.
For that we start with results from studies of PPDs around low mass stars, and our
assumptions of similarity to VMS.
We impose four conditions for fragmentation objects to be formed around VMSs:

(1) Planets are formed by fragmentation in the PPD where the surface density is
$\sim 10-100 \rm{~g~cm^{-2}}$ (e.g., Nero \& Bjorkman 2009).
We assume that fragmentation in PPDs around VMSs occurs at the same surface density.

(2) The solar PPD mass was $\sim 1-20$ per cent of the mass of the Sun
(Weidenschilling 1977; Hayashi 1981; Kuchner 2004; Desch 2007).
Again, we take approximately the same fraction to hold in our study, which translates to a total PPD mass
of $\sim 1-20 \rm{M_{\odot}}$.
This mass leads to a relation between the surface density and the outer radius of the PPD, $R_d$
\begin{equation}
m_d = \int_0^{R_d} \Sigma(r)2 \pi r \,dr,
\label{eq:mdisk1}
\end{equation}
where we assume that the inner radius of the PPD is small and can be neglected
(for $p < 2$ in equation (\ref{eq:sigma1})).

(3) In order for the PPD to fragment, two conditions have to be satisfied (e.g., Kratter et al. 2010).
(3.1) Toomre's parameter (Toomre 1964) should satisfy
\begin{equation}
Q=\frac{c_s \Omega}{\pi G \Sigma} < Q_f \sim 1,
\label{eq:toomre}
\end{equation}
where $\Omega$ is the Keplerian angular velocity, $c_s=\sqrt{kT/ \mu m_p}$ is the speed of sound,
$m_p$ is the proton mass, $\mu$ is the molecular weight, and $k$ is the Boltzmann constant.
(3.2) The radiative cooling time (in the optically thick PPD) must be shorter than the orbital time (Gammie 2001),
\begin{equation}
t_{\rm{cool}} \lesssim \frac{3}{\Omega}
\label{eq:tcool}
\end{equation}
so the fragment can cool in less than about one orbital period.
According to the model of Kratter et al. (2010) there is a minimal orbital distance which satisfies
both conditions for fragmentation , the fragmentation distance.
For a $1.5 \rm{M_{\odot}}$ star, Kratter et al. (2010) found this radius to be $\sim 70 \AU$.
An approximatively similar fragmentation distance was obtained by Clarke (2009).
Based on the results of Kratter et al. (2010) and our assumptions, we scale with a VMS to derive
the fragmentation distance appropriate for our studied parameter space
\begin{equation}
r_f \simeq 280 \left(\frac{M_*}{100 \rm{M_{\odot}}}\right)^{\frac{1}{3}} \AU. 
\label{eq:rfrag}
\end{equation}
The same fragmentation distance can also be determined from the condition that the vertical
optical depth is $\tau \sim 1$ (Kratter et al. 2010).
As can be seen the fragmentation distance weakly depends on the mass of the central star.
As for VMSs the mass-luminosity relation is $M_* \propto L_*^\beta$, where $\beta \simeq 0.3$,
so the fragmentation distance very weakly depends on the luminosity, $r_f \propto L_*^{\frac{\beta}{3}}$.

(4) Observations show that planets form quite close to the fragmentation distance.
For example, Fomalhaut b, the outer planets in the triple-planetary system HR 8799, and the potential
protoplanet associated with HL Tau, were all formed outer to, but quite close to the fragmentation radius,
at distances of $\sim (1-3) r_f$.
For our studied cases we approximate this range as $ \sim (1-5) r_f$
which translates to $\sim 300-2000 \AU$ if we consider VMSs with masses up to $\sim 300 \rm{M_{\odot}}$.
We would therefore expect to find the inner fragmentation objects approximately within this range.
Outer fragmentation objects can reside at larger orbital distances, up to $\sim 3000 \AU$.

We build three calibrations which obey all the conditions listed above, with the main difference
being the power $p$ of the surface density profile given in equation (\ref{eq:sigma1}).
First, we examine the implications of the demand that for $M_* = 100 \rm{M_{\odot}}$
the fragmentation object is formed at $r_p=1000\AU$ and that the surface density
there is the minimal value allowed by condition (1), $\Sigma_p = 10 \rm{~g~cm^{-2}}$.
We therefore obtain our three density profiles
\begin{equation}
\Sigma(r) \simeq
  \begin{cases}
10 \left(\frac{r}{1000 \AU}\right)^{-1}
 \left(\frac{M_*}{100 \rm{M_{\odot}}}\right)^{\frac{1}{3}} \rm{~g~cm^{-2}} &; p=1 
 \\
10 \left(\frac{r}{1000 \AU}\right)^{-\frac{5}{4}}
 \left(\frac{M_*}{100 \rm{M_{\odot}}}\right)^{\frac{5}{12}} \rm{~g~cm^{-2}} &; p=\frac{5}{4} 
 \\
10 \left(\frac{r}{1000 \AU}\right)^{-\frac{3}{2}}
\left(\frac{M_*}{100 \rm{M_{\odot}}}\right)^{\frac{1}{2}} \rm{~g~cm^{-2}}  &; p=\frac{3}{2}. 
  \end{cases}
\label{eq:sigma2}
\end{equation}
As can be seen from equation (\ref{eq:sigma2}) and the four conditions,
there is some freedom in the values of the different numerical factors,
and the set of numbers is not unique.
However, the possible range of numerical factors do not change much our conclusions.
The ratio between the orbital distances $r_p$ where fragmentation around VMSs occurs and the
fragmentation distance $r_f$ is taken to be constant $\chi=r_p/r_f$.
This constant ratio is set such that for $M_* = 100 \rm{M_{\odot}}$ we shall get $r_p = 1000 \AU$.
As can be seen from equation (\ref{eq:rfrag}) this ratio is $\chi \simeq 3.5$. 
We therefore get, for all three calibrations
\begin{equation}
r_p \simeq 1000 \left(\frac{M_*}{100 \rm{M_{\odot}}}\right)^{\frac{1}{3}} \AU.
\label{eq:rp1}
\end{equation}
Note that according to our conditions (3) and (4) the fragmentation objects can be formed at larger
distanced than $r_p = 1000 \AU$, up to
\begin{equation}
r_{p,\rm{max}} \simeq 1500 \left(\frac{M_*}{100 \rm{M_{\odot}}}\right)^{\frac{1}{3}} \AU. 
\label{eq:rpmax}
\end{equation}
The PPD is assumed to posses a total mass of $m_d=0.2M_*$, the maximum allowed by condition (2),
and therefore extends up to
\begin{equation}
R_d \simeq
  \begin{cases}
2800 \left(\frac{M_*}{100 \rm{M_{\odot}}}\right)^\frac{2}{3} \AU &; p=1           
 \\
2700 \left(\frac{M_*}{100 \rm{M_{\odot}}}\right)^\frac{7}{9} \AU &; p=\frac{5}{4} 
 \\
2100 \left(\frac{M_*}{100 \rm{M_{\odot}}}\right) \AU             &; p=\frac{3}{2}. 
  \end{cases}
\label{eq:rdisk1}
\end{equation}

The most simple approximation for the mass of the fragmentation object,
using the models of Goodman \& Tan (2004) and Kratter \& Matzner (2006) gives
\begin{equation}
\begin{split}
m_f & = \frac{4\pi}{\Omega_p} \frac{c_{s,p}^3}{G} Q_p\\
    & \simeq 0.06 \left(\frac{Q_f}{1}\right)^4
\left(\frac{\Sigma_p}{10 \rm{~g~cm^{-2}}}\right)^3
\left(\frac{r_p}{1000 \AU}\right)^{-4p}
\rm{M_{\odot}}, 
\end{split}
\label{eq:mfrag4}
\end{equation}
where $\Sigma_p \equiv \Sigma(r_p)$ was calibrated to a range compatible with planet formation
as discussed above, $\Omega_p \equiv \Omega(r_p)$, $Q_p \equiv Q(r_p)$ and $c_{s,p} \equiv c_s(r_p)$.
In the second equality of equation (\ref{eq:mfrag4}) we used equations ($\ref{eq:toomre}$), ($\ref{eq:rp1}$),
and the expression for the speed of sound.

We will also check an approximation from another model.
As the PPD is assumed to be optically thick at the distance where it fragments,
the mass of the fragmentation object can be approximated as in Nero \& Bjorkman (2009)
\begin{equation}
m_f \simeq 0.1 \left(\frac{\Sigma_p}{10 \rm{~g~cm^{-2}}}\right)^{\frac{3}{2}}
\left(\frac{r_p}{1000 \AU}\right)^{3}
\left(\frac{M_*}{100 \rm{M_{\odot}}}\right)^{-\frac{1}{2}} \rm{M_{\odot}}. 
\label{eq:mfrag2}
\end{equation}

Note that in equations (\ref{eq:mfrag4}) and (\ref{eq:mfrag2}) $\Sigma_p$ and $r_p$ are related
by equation (\ref{eq:sigma2}) and cannot be individually calibrated.
Substituting equations (\ref{eq:sigma2}) and (\ref{eq:rp1}) in equation (\ref{eq:mfrag2}) we get
\begin{equation}
m_f \simeq 0.1 \left(\frac{M_*}{100 \rm{M_{\odot}}}\right)^{\frac{3}{2}} \rm{M_{\odot}},
\label{eq:mfrag3}
\end{equation}
namely the dependence of $m_f$ in the mass of the VMS is the same for our three calibrations.
Note that the typical mass of the fragmentation object is on the boundary of BDs
and low mass stars.

Let us estimate the temperature at the radius $r_p$ where fragmentation takes place.
The simplest estimate is to use Toomre's parameter (equation (\ref{eq:toomre})), and to substitute
the explicit expression for the sound speed.
This gives for the temperature
\begin{equation}
\begin{split}
T(r_p) &= \frac{\pi^2 \mu m_p G^2 Q_f^2 \Sigma_p^2}{k \Omega_p^2}\\
       &\simeq 20 \left(\frac{\mu}{1.3}\right)
\left(\frac{Q_f}{1}\right)^2
\left(\frac{r_p}{1000 \AU}\right)^{-2p}
\K. 
\end{split}
\label{eq:Tcrp0}
\end{equation}
where in the second equality we used equation (\ref{eq:rp1}) to eliminate the dependance on
the stellar mass (which comes from $\Omega_p$).

We will make another estimate of the temperature ar $r_p$.
Using the $\alpha$-disk model for PPDs as in Dullemond et al. (2007),
we can estimate the intrinsic temperature at the center of the PPD which depends on the radius as
$T_c(r) \propto r^{-\frac{1}{2}}$.
The PPD is assumed to be geometrically thin, and the temperature is assumed to result from the
internal viscosity of the PPD, and not from the radiation of the parent VMS.
Namely, the PPD is horizontally optically thick.
The model of Dullemond et al. (2007) uses, as in our second calibration, a surface density profile of
$\Sigma \propto r^{-1}$, explicitly written as
\begin{equation}
\Sigma(r) = \frac{\sqrt{G M_*} \mu m_p \dot{M}}{3 \pi k \alpha T_c(r) r^{\frac{3}{2}}},
\label{eq:sigma4}
\end{equation}
where $\dot{M}$ is the mass accretion rate of the PPD and $\alpha$ is the $\alpha$-disk parameter (Shakura \& Sunyaev 1973).
Vaidya et al. (2009) found that $\alpha \sim 0.1$ is an optimal value in order for a PPD to be
on one hand stable against complete fragmentation due to thermal effects caused by high viscosity (occurs for $\alpha \gtrsim 1$),
and on the other hand not completely subjected to gravitational instability (occurs for $\alpha \lesssim 0.01$).
Namely, setting $\alpha \sim 0.1$ well describes the situation observed in PPDs, where fragmentation
occurs beyond some fragmentation distance, such as given in equation (\ref{eq:rfrag}).
For detailed analytical calculations of $\alpha$ and the PPD's viscosity profile see Isella et al. (2009).

Equating equation (\ref{eq:sigma4}) with the first calibration of equation (\ref{eq:sigma2}),
we find that the PPD's central (midplane) temperaturetemperature in the radius $r_p$ where fragmentation takes place is
\begin{equation}
\begin{split}
T_c(r_p) &\simeq 70
\left(\frac{\alpha}{0.1}\right)^{-1}
\left(\frac{\mu}{1.3}\right)
\left(\frac{\dot{M}}{10^{-5} \rm{M_{\odot}~yr^{-1}}}\right)\\
         &\left(\frac{\Sigma_p}{10 \rm{~g~cm^{-2}}}\right)^{-1}
\left(\frac{r_p}{1000 \AU}\right)^{-\frac{3}{2}}
\left(\frac{M_*}{100 \rm{M_{\odot}}}\right)^{\frac{1}{2}}
\K \\
         &\simeq 70
\left(\frac{\alpha}{0.1}\right)^{-1}
\left(\frac{\mu}{1.3}\right)
\left(\frac{\dot{M}}{10^{-5} \rm{M_{\odot}~yr^{-1}}}\right)
\K.
\end{split}
\label{eq:Tcrp1}
\end{equation}
In the second equality we used equations (\ref{eq:rp1}) and the first calibration in equation (\ref{eq:sigma2})
to eliminate the dependance on the $\Sigma_p$, $r_p$, and $M_*$.

We consider also the case were the central star radiation is not blocked but rather heat the disk at $r_p$.
An estimate from another model, by Dodson-Robinson et al. (2009), based on an older model
of Chiang \& Goldreich (1997) which takes the existence of dust into account, gives
(when translated from stellar radius and effective temperature to luminosity)
\begin{equation}
T(r_p) \approx
200 \left(\frac{\theta}{0.1}\right)^{\frac{1}{4}}
\left(\frac{r_p}{1000 \AU}\right)^{-\frac{1}{2}}
\left(\frac{L_*}{2\times10^6 \rm{L_{\odot}}}\right)^{\frac{1}{4}}
\K, 
\label{eq:Tcrp2}
\end{equation}
where $L_*$ is luminosity of the VMS, and $\theta$ is the flaring angle of the PPD at $r=r_p$.

The temperature estimates in equations (\ref{eq:Tcrp0}) and (\ref{eq:Tcrp0})
are compatible with estimates of temperatures where fragmentation
usually takes place, and planets are formed $\lesssim 100 \K$
(e.g., Boss 1998, 2009; Dodson-Robinson et al. 2009).
The high temperature derived in equation (\ref{eq:Tcrp2}) shows that if the radiation from the VMS is not blocked,
then $Q \gg 1$, and there will be no fragmentation.

Our most significant conclusion of this section is that in PPDs around VMS, low mass stars
and BDs can be formed in the same way as planets are formed around solar-like stars.
For example, we might find 2 or more low mass stars and BDs orbiting the parent star in the same plane
as BDs and planets around low mass stars.

Though in some cases our calculation can be refereed to as guidelines, and
the exact numbers might be somewhat different,
we showed that using different approaches to calculate the physical
properties of the PPD, we find that it is likely to fragment and
form an object in the PPD.

\section{OBSERVATIONAL PREDICTIONS}
\label{sec:obs}

The ability to observe fragmentation objects depends strongly on their masses,
and their orbital distance from the parent VMS.
The fragmentation objects to parent star typical mass ratio is very small, $\sim 10^{-3}$,
and their typical orbital period is extremely long, $\gtrsim {\rm few} \times 1000 \yrs$.
It is impossible to detect such objects by the Doppler shift method.
If the fragmentation object is a planet it would only reflect the light of its parent star,
while if it is a BD or a star it would produce its own
luminosity, peaked in the infrared.

\begin{figure}[!t]
\resizebox{0.5\textwidth}{!}{\includegraphics{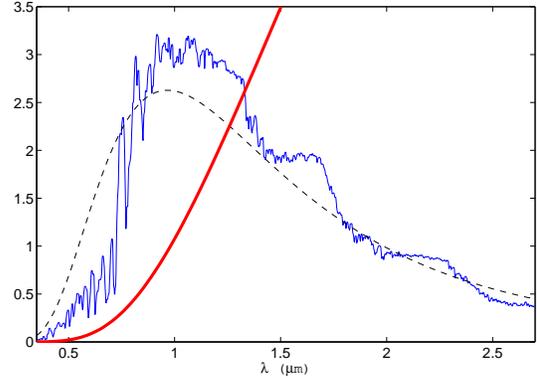}}
\caption{\footnotesize
Dashed thin black line: theoretical blackbody flux (in units of $10^{26} \rm{ \erg~s^{-1} {\AA}^{-1}}$)
from a low mass star or brown dwarf with an effective temperature of $3000\K$ and a luminosity of
$10^{-3} \rm{L_\odot}$ in an early evolution stage.
Solid thin blue line: theoretical model (Rice et al. 2010) for emission of a $3000\K$ low mass star
(in units of $10^{26} \rm{ \erg~s^{-1} {\AA}^{-1}}$).
Solid thick red line: The flux ratio ($\times 10^{-7}$) between a ($3000\K$; $10^{-3} \rm{L_\odot}$) low mass star or
brown dwarf and a ($50000\K$; $2 \times 10^6 \rm{L_\odot}$) very massive star, both assumed to emit blackbody radiation.
The James Webb Space Telescope (JWST) will observe in the band
$\lambda = 0.6 - 26 \mum$, and can detect the fragmentation objects studied here.
}
\label{fig:JWST}
\end{figure}

In a relatively short time a VMS enters the Luminous Blue Variable and Wolf-Rayet stages of evolution,
in which it expels a considerable fraction of its mass (e.g., Smith \& Owocki 2006).
Suppose it expels a fraction $\Delta M / M_*$ of its mass over a time scale longer than the
orbital period.
The orbital radius of the fragmentation object would move to a larger distance of
\begin{equation}
r_{p,2} \simeq 5000 \left( \frac{r_p}{1000 \AU} \right) \left( \frac{M_*}{100 \rm{M_{\odot}}} \right)
\left( \frac{M_* - \Delta M}{20 \rm{M_{\odot}}} \right)^{-1} \AU,
\label{eq:rp2}
\end{equation}
where $r_p$ is the orbital distance where the fragmentation object is formed (Eggleton 2006).
If there are short phases of large mass loss the orbit will
become eccentric, with the distance to apastron, where the companion
spends most of the time, larger even.
Low mass stellar companions might be photometrically detected at large orbital separations of
$\sim 2000 - 10^4 \AU$, as the separation at a distance of, say, $2 \kpc$ be $\sim 1-5^{\prime \prime}$.
For example, it might be possible to detect a low mass star with the James Webb Space Telescope (JWST).
We take a low mass star (or a BD) with an effective temperature of $3000 \K$
and luminosity of the order of $\sim 10^{-3} \rm{L_\odot}$,
and a VMS with an effective temperature of $5\times10^4\K$ and luminosity of $\sim 2 \times 10^6 \rm{L_\odot}$,
and find that the ratio between the fluxes of the two stars at $\lambda=1 \mum$, close to the wavelength
where the fragmentation object peaks, is $\sim 10^{-7}$. 
As the separation is large, this ratio should be detected by the JWST.
This is depicted in Fig. \ref{fig:JWST}.

In the scenario described in section \ref{sec:frag}, more than one fragmentation object
may be formed.
If a few planets, BDs or low mass stars are formed, they are expected to
(1) be located in the same orbital plane, where the PPD used to lie,
(2) have circumstellar distances in the range of a $\sim {\rm few} \times 100 - 10^4 \AU$,
and
(3) have a total mass of a few per cents of the parent VMS.
We predict that observations of VMSs with the JWST and other telescopes of high capabilities in the
IR bands can reveal systems where these conditions are fulfilled.

We note the dramatic difference in the evolutionary timescale
of the parent star and the fragmentation objects.
A VMS is expected to go through its entire evolutionary path in only a ${\rm few} \times 10^6 \yrs$
(e.g., Maeder \& Meynet 1994).
But a low mass star would not even reach its zero-age main sequence by then
(e.g., D'Antona \& Mazzitelli 1994), neither a BD would start burning Deuterium
(e.g., D'Antona \& Mazzitelli 1985).
At the age of ${\rm few} \times 10^6 \yrs$ the temperature of developing BDs may be
$\sim 2000-5000 \K$ (e.g., Chabrier et al. 2000), approximately in the same range as for low mass stars.

The estimations of Kratter \& Matzner (2006) suggest that the disk-born stars
(with masses of $\sim 1 \rm{M_{\odot}}$ for $M_* = 100 \rm{M_{\odot}}$) are formed near
or somewhat outside $r_p \simeq 100-200 \AU$ and therefore there is a very low chance to detect them.
According to our estimate it is possible that the fragmentation objects reach up to
$\sim 10^4 \AU$, and thus we predict that these low mass stars and BDs can be
relatively easily observed.

Recent observations of Smith et al. (2009) reveal an intriguing potential
candidate for a fragmentation object resembling the one we suggest.
The source G in W49A appears to be a $45 \rm{M_\odot}$ protostar surrounded by a disk
from which it was formed.
The PPD is extending up to a distance of $\sim 6 \times 10^4 \AU$, 
and the inner disc has been probably cleared out to a radius of $\sim 1000 - 2000 \AU$.
Within the disk, in the same plane, there appears to be a second star located $\sim 4 \times 10^4 \AU$ 
from the central source.
Though these distances are larger by about one order of magnitude than our preliminary calibration,
the common orbital plane may be an indication that this star formed in the disk,
perhaps in fragmentation process as we suggest.

Vlemmings et al. (2010) have discovered a strong magnetic field around the massive protostar Cepheus A HW2,
which also has a disk with a radius of $\sim600 \AU$ surrounding it. 
This discovery suggests that strong magnetic fields threading the PPD and maintaining a high accretion rate to the star,
which is essential to its formation.
It is not yet clear to us if these magnetic fields are helpful for planet formation as though they assist the
formation of the star, they may as well moderate instabilities in the disk and make fragmentation more difficult.

\section{PHOTOEVAPORATION CONSIDERATIONS}
\label{sec:photoevaporation}

As we suggest that even low mass stars can be formed around VMSs, a point worth discussing
is whether the radiation of low mass stars prevent their creation.
Our answer to this question is most likely no.
The track on the HR diagram in which low mass stars settle in the main sequence,
starts in luminosity much higher than the luminosity the low mass stars have when they
reach the main sequence.
Thus, as the PPD fragments the low mass star has already passed the critical point in which its
luminosity had the chance to prevent its creation.
We therefore do not expect the radiation of the fragmentation object be an obstacle for the
low mass star formation.

Krumholz et al. (2009) performed 3D hydrodynamical simulations of the formation
of a massive star via accretion from a disk.
It was clearly shown that during the formation of a massive star, while it is still accreting through a PPD,
the radiation it emits does not prevent accretion, and neither causes the PPD's destruction.
Most of the radiation of the massive star is instead tunneled into polar directions,
where it creates two low density polar bubbles.
Material from the cloud continues to be accreted onto the star, and when it comes from polar directions it flows
along the walls of the bubbles and accreted onto the star.
The PPD remains for a few~$\times 10^4 \yr$, long enough for fragmentation to occur, and form a companion star in the disk.
Kuiper et al. (2010) have recently developed a promising frequency dependent radiation transport simulation
that is likely to strengthen the results of Krumholz et al. (2009).

\begin{figure}[!t]
\resizebox{0.5\textwidth}{!}{\includegraphics{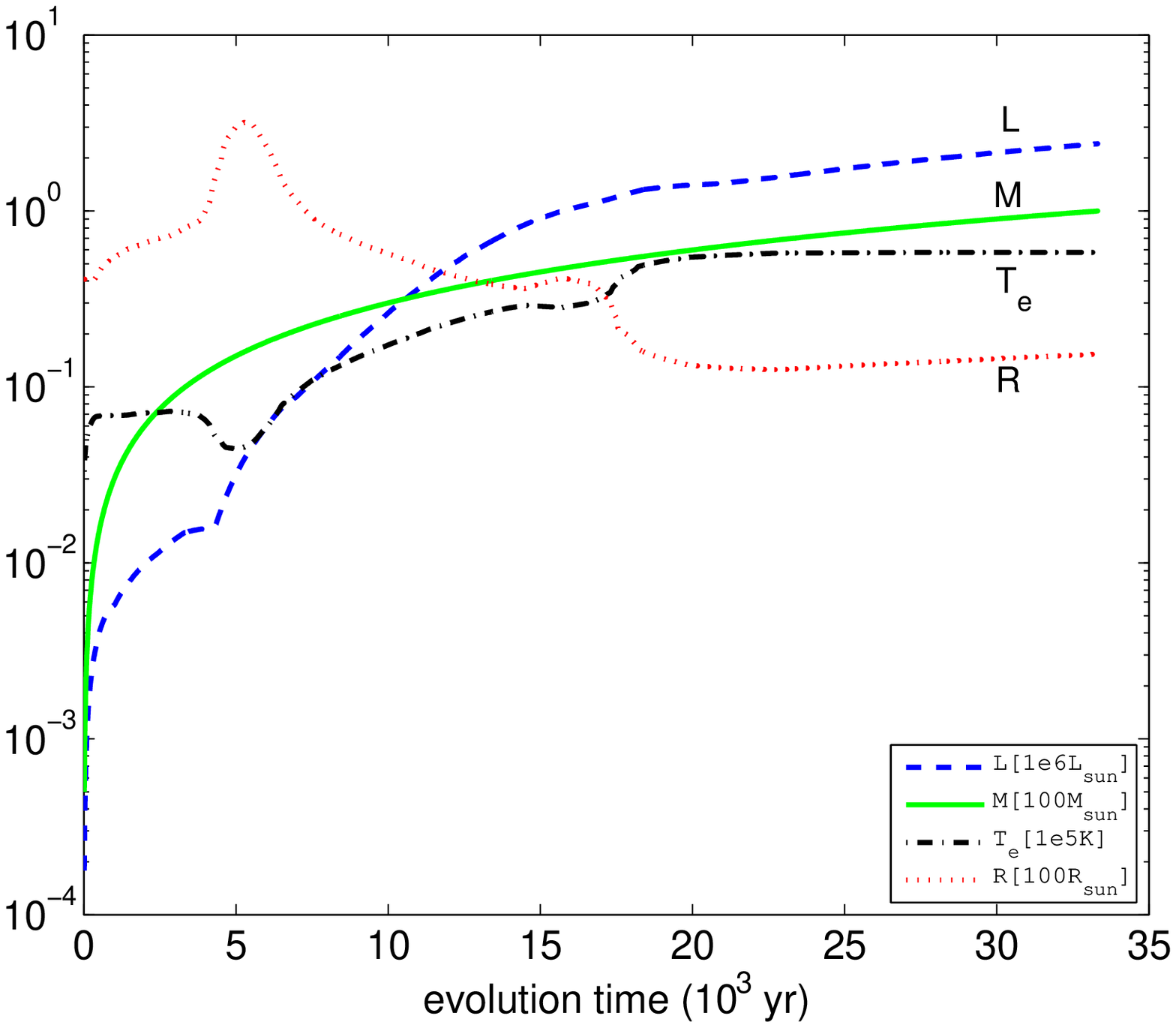}}
\resizebox{0.5\textwidth}{!}{\includegraphics{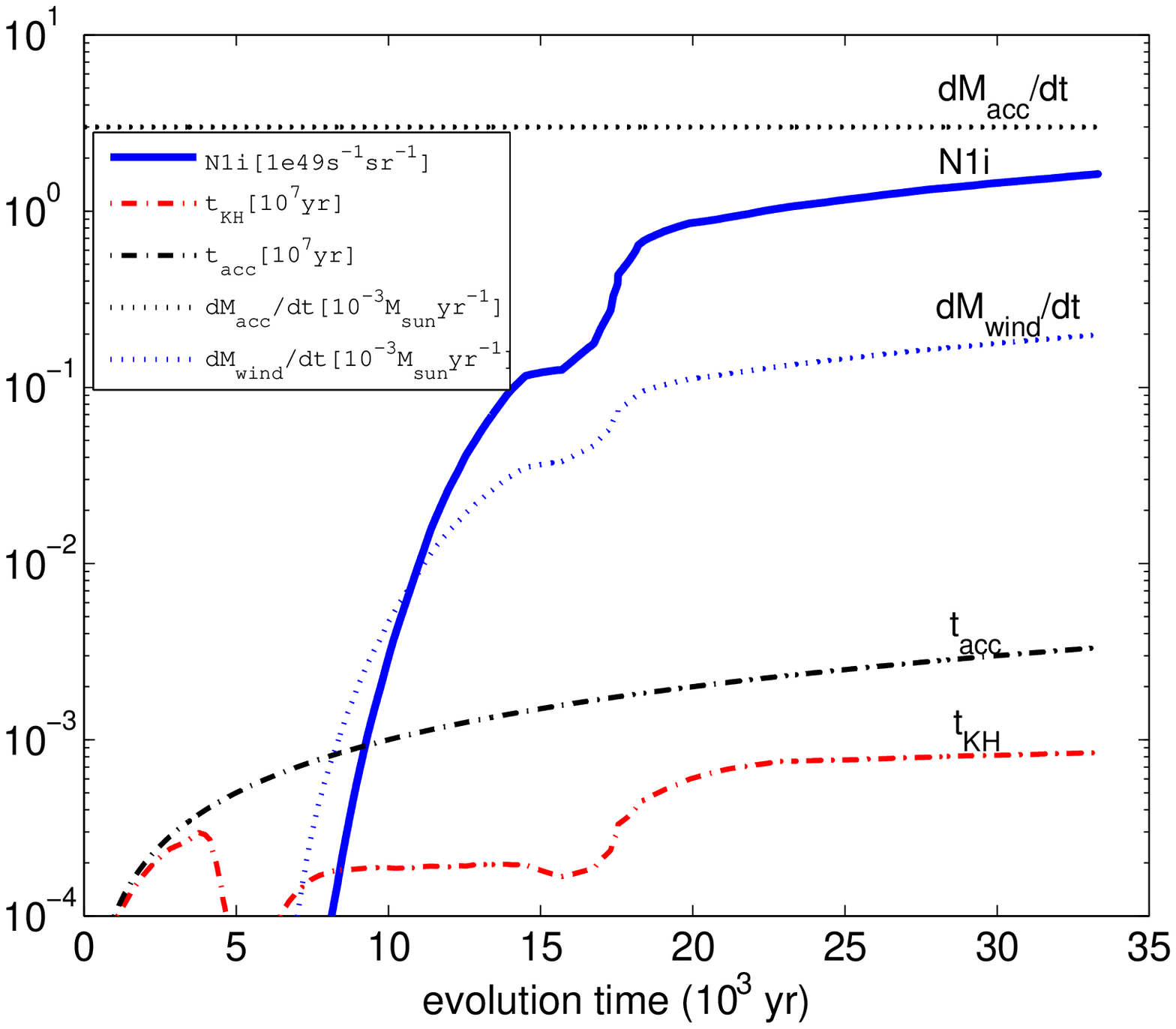}}
\caption{\footnotesize
First panel: The luminosity, mass, effective temperature and radius of the star as a function of its evolution time,
as inferred from the model of Hosokawa \& Kazuyuki (2009).
Second panel: Calculated quantities based on the model of Hosokawa \& Kazuyuki (2009): Mass accretion rate,
the Kelvin-Helmholtz time, accretion time from the disk, ionizing photons rate, and mass loss rate by the disk wind
(eq. \ref{eq:photoevaporation1}).
When integrating the mass loss rate by disk wind over the evolution time (equation (\ref{eq:photoevaporation2}) we find that
only $\sim 2.5 \rm{M_\odot}$ were lost from the PPD, and therefore it survived photoevaporation.
}
\label{fig:photoevaporation}
\end{figure}

Radiation from massive star photo-evaporates their PPD, and lead to the formation of a disk wind.
We will examine the degree to which photoevaporation by VMS radiation disrupt
the PPD and prevents fragmentation.
We use the results from Hosokawa \& Kazuyuki (2009) who modeled the evolution of a newly formed accreting star
reaching a final mass of $100 \rm{M_\odot}$ during its formation
(their model MD3$\times$3, presented in their figure 18).
In this model of Hosokawa \& Kazuyuki (2009) the mass accretion rate is fixed at
$M_{\rm{acc}}=3\times 10^{-3} \rm{M_\odot yr^{-1}}$,
and the VMS reaches its final mass in $\sim 3.3 \times 10^4 \yr$.
From the model of Hosokawa \& Kazuyuki (2009) we calculate all the basic parameters of the star
as a function of its evolution time.
The first panel of Figure \ref{fig:photoevaporation} shows the luminosity, mass,
effective temperature and radius of the star as a function of its evolution time.
The final luminosity of the star is $T_e = 58,200K$,
and its effective temperature is $L = 2.41 \times 10^4 \rm{L_\odot}$.
Assuming a blackbody radiation, we calculate the ionizing photon rate for this model
(the results of Schaerer \& de Koter 1997 show that this is an adequate approximation).
At $t = 1.7\times10^4 \yr$ the ionizing photons rate rapidly raises,
to $4\pi N_{i1}\approx 10^{50} \rm{s^{-1}}$ and then disk photoevaporation should be considered.

We follow the model of Hollenbach et al. (1994) which related the ionizing photons rate to the
mass loss rate expected from the PPD
\begin{equation}
\dot{M}_{\rm{wind}} \simeq 1.4 \times 10^{-4}
\left(\frac{4\pi N_{i1}}{10^{50} \rm{s^{-1}}}\right)^{\frac{1}{2}}
\left(\frac{M}{100 \rm{M_\odot}}\right)^{\frac{1}{2}}. 
\label{eq:photoevaporation1}
\end{equation}
More detailed models (e.g. Font et al. 2004; Alexander et al. 2006) give similar values to those obtained by
Hollenbach et al. (1994).
The second panel of Figure \ref{fig:photoevaporation} shows the ionizing photons rate, the disk wind mass loss rate
as well as some other quantities.

To find the total mass lost from the PPD by photoevaporation we integrate the mass loss rate over the
evolution time of $\sim 3.3 \times 10^4 \yr$ and find
\begin{equation}
M_{\rm{lost}} =  \int \dot{M}_{\rm{wind}}~{\rm d}t  \simeq 2.5 \rm{M_\odot}. 
\label{eq:photoevaporation2}
\end{equation}
This is much smaller than the mass of the PPD in our model, and can be regarded as a negligible loss of mass.
We therefore conclude that photoevaporation cannot destroy the disk during the relevant time for fragmentation.

\section{SUMMARY AND DISCUSSION}
\label{sec:summary}

Planets are found in a variety of orbital distances, around stars ranging in mass from low mass to massive,
in both metal rich and metal poor environments, around single stars and binaries, and around
stars in different stages of evolution (see the Extrasolar Planets Encyclopaedia,
maintained by Schneider 2010, and references therein;
for a recent statistical analysis see Mordasini et al. 2009).
In the present paper we suggested that very massive stars (VMSs) of $M_* \gtrsim 100 \rm{M_{\odot}}$
are very likely to also harbor planetary systems.

Under the assumption that fragmentation of protoplanetary disk (PPD) around solar like stars
can be scaled to match the properties of PPD around VMS, we have found here that it is likely
that in such PPDs, BDs and low mass stars of $\sim 0.1-0.3 \rm{M_{\odot}}$ are likely to be formed.
We suggest that these fragmentation objects
are orbiting the VMS in circumstellar distances in the range of
$\sim {\rm few} \times 100 - 10^4 \AU$, on the same orbital plane.
The low mass stars might be photometrically detected in the near infrared with modern ground
telescopes using adaptive optics in the IR, and modern space telescopes such as the JWST.

As in low mass stars (e.g., Qian et al. 2010) it is possible that the massive fragmentation objects
will be formed around binary systems of VMS in P-type orbits (outside the two components of the binary system).
The fragmentation objects have very large orbital distances that ensure
stability (e.g., Szebehely 1980; Dvorak et al. 1989; Musielak et al. 2005).
We note that simulations by Krumholz et al. (2009) have shown that fragmentation can be possible
also if in the center of the PPD lie a massive binary system.
In the same simulation, the massive binary companion itself
was formed from fragmentation objects in the PPD around the primary which merged into a massive stellar companion.

Theoretical calculations based on the common stelar formation process limit the
mass of the newly born star to $M \la 150 M_\odot$ (e.g., Figer 2005).
However, it seems that stars can reach higher masses, e.g., the Pistol star
(Figer et al. 2004; Najarro 2005)) and $\eta$ Car (Kashi \& Soker 2010).
These higher masses suggest that accretion from the PPD continues for a relatively longer time,
with lots of mass processed through the PPD.
Merger of a number of stars to form VMS also implies that there is lots of material in the
surrounding of the newly formed star.
The PPD itself is replenished by the surrounding gas cloud.
Our study shows that such PPD are likely to leave fragmentation objects and debris
(smaller objects and dust).

Our calculations focused on large fragmentation objects.
Around solar-like stars these are planets, while around VMS these might be scaled to
brown dwarfs and low mass stars.
We can take another, somewhat speculative, step, and scale the Kuiper belt from
the solar system to VMS.
In VMSs, such objects might be $\sim 100$ times as massive as those in the
solar system, and Kuiper objects are scaled to Mercury like planets.

It is possible that such planets will sustain water in its liquid state.
Crudely extrapolating the conditions for stable P-type orbits
(e.g., Musielak et al. 2005) to a binary mass ratio of $M_2/M1=10^{-3}$
implies that a planet can have a stable orbit at $a \ga 1.6$ times the binary separation.
Let us consider a VMS with a luminosity of $L_* \sim 2-5\times10^6 \rm{L_{\odot}}$,
for which a temperature similar to that on earth will be found at a circumstellar
distance of $a \sim 1500-2200 \AU$ (depending on atmospheric conditions).
Therefore, if the outer large fragment object (of mass $\sim 0.1-0.2 M_\odot$) is at
$a \la 1500 AU$, Mercury type planets might exist in an habitable zone.
such a zone around VMS might contain hundreds of Mercuty like planets,
and many more smaller objects.
Collisions between these will form dust.
Therefore, a zone of excess in IR emission might exist there.
In a future paper we will examine ways to detect debris disks around VMSs.

The PPDs around VMSs are expected to be huge and very massive relative to those around
solar-like stars.
These types of disks have favorable conditions for formation circumstellar
objects via fragmentation, rather than core accretion.
Population III stars are thought to be of VMSs.
If our results hold for very low metallicity populations of VMSs, then it is not
unreasonable to consider the formation of planets and brown dwarfs (and of course low mass stars)
even around population III stars.
Probably not around the very first stars, but around those that contain a very low abundance of metals,
i.e., late population III stars with metallicity of $\sim 10^{-6}$ times solar,
and early population II stars.
The VMSs explode eventually, and the fragmentation objects become unbound (if more than half the mass is
lost in the explosion).
We therefore raise the possibility that the (rare) most metal-poor stars in the galaxy were
formed in PPDs of the very first Population II VMSs.

We thank
Scott Kenyon, 
Kaitlin Kratter, 
Ken Rice, 
Nathan Smith, 
Ralph Neuh\"{a}user, 
Dimitris Stamatellos, 
and
Bhargav Vaidya 
for helpful comments.
This research was supported by the Asher Fund for Space Research at the
Technion, and the Israel Science Foundation.

\end{document}